\documentclass{tcibook}
\usepackage{fancyhea}
\usepackage{work}
\usepackage{bm}       \usepackage{graphicx}
\usepackage{multirow}
\usepackage{lineno}
\usepackage{threeparttable}
\usepackage[normalem]{ulem}
\usepackage{color}
\usepackage[colorlinks = true,
            linkcolor = blue,
            urlcolor  = blue,
            citecolor = blue,
            anchorcolor = blue]{hyperref}
\usepackage{amsmath,amssymb}

\def\beq{\begin{equation}}
\def\eeq#1{\label{#1}\end{equation}}
\def\eeqn{\end{equation}}

\newenvironment{Eqnarray}{\arraycolsep 0.14em\begin{eqnarray}}{\end{eqnarray}}
\def\beqa{\begin{Eqnarray}}
\def\eeqa#1{\label{#1}\end{Eqnarray}}
\def\eeqan{\end{Eqnarray}}

\let\bar=\overbar

\def\lsim{\mathrel{\raise.3ex\hbox{$<$\kern-.75em\lower1ex\hbox{$\sim$}}}}
\def\gsim{\mathrel{\raise.3ex\hbox{$>$\kern-.75em\lower1ex\hbox{$\sim$}}}}

\def\del{\partial}
\def\Dslash{\not{\hbox{\kern-4pt $D$}}}
\def\dslash{\not{\hbox{\kern-2pt $\del$}}}
\def\pslash{\not{\hbox{\kern-2pt $p$}}}
\def\ETmiss{\not{\hbox{\kern-4pt $E$}}_T}

\def\Dlr{\mathrel{\raise1.5ex\hbox{$\leftrightarrow$\kern-1em\lower1.5ex\hbox{$D$}}}}

\def\BR{\mbox{\rm BR}}

\def\MSB{{\bar{M \kern -2pt S}}}
\def\msb{{\bar{\scriptsize M \kern -1pt S}}}

\def\drb{{\bar{\scriptsize D \kern -1pt R}}}

\def\authorlist#1#2{
    \vskip 0.4in
\begin{center}\begin{large} {\bf  #1 } \end{large}
    \vskip 0.2in
              #2
     \vskip 0.2in
   \end{center}
}

\usepackage{xspace}
\newcommand{\gm}{\ensuremath{g\!-\!2}\xspace}

\begin{document}

\pagenumbering{roman}

\parindent=0pt
\parskip=8pt
\setlength{\evensidemargin}{0pt}
\setlength{\oddsidemargin}{0pt}
\setlength{\marginparsep}{0.0in}
\setlength{\marginparwidth}{0.0in}
\marginparpush=0pt

\pagenumbering{arabic}

\renewcommand{\chapname}{chap:intro_}
\renewcommand{\chapterdir}{.}
\renewcommand{\arraystretch}{1.25}
\addtolength{\arraycolsep}{-3pt}

\setcounter{chapter}{2}

\chapter{Fundamental Physics in Small Experiments}

\authorlist{T. Blum, P. Winter}
   {T.~Bhattacharya, T.Y.~Chen, V.~Cirigliano, D.~DeMille, A.~Geraci, N.R.~Hutzler, T.M.~Ito, D.~Kaplan, O.~Kim, R.~Lehnert, W.M.~Morse, Y.K.~Semertzidis}

\section{Introduction}
High energy physics aims to understand the fundamental laws of particles and their interactions at both the largest and smallest scales of the universe. This typically means probing very high energies or large distances or using high-intensity beams, which often requires large-scale experiments on earth or in space. Relevant projects for this approach, like particle colliders, telescopes, or other large-volume detectors have provided tremendous insights but come with increasingly large associated costs. 

A complementary approach to probe high energy scales is offered through high-precision measurements in small- and mid-scale size  experiments, often at lower energies. The field of such high-precision experiments has seen tremendous progress and importance for particle physics for at least two reasons. First, they exploit synergies to adjacent areas of particle physics and benefit by many recent advances in experimental techniques and  technologies. Together with intensified phenomenological explorations, these advances have led to the realization that challenges associated with weak couplings or the expected suppression factors from the mass scale of new physics can be overcome with such methods while maintaining a large degree of experimental control. Second, many of these measurements add a new set of particle physics phenomena and observables that can be reached compared to the more conventional methodologies using high energies. Combining high-precision measurements of smaller-scale experiments with the large-scale efforts described above therefore casts both a wider and tighter net for possible effects originating from physics beyond the Standard Model (BSM).

The efforts described in this section are focused on providing crucial insights into possible BSM physics by studying areas such as novel short-range interactions, the small-scale structure of spacetime and in particular the fate of Lorentz, translation, CPT, CP, T, and P symmetries, the gravitational interaction of antimatter, certain quantum aspects of gravity, millicharged particles, gravitational-wave measurements, dark matter and dark energy. The growing impact of these high-precision studies in high energy physics and the complementary input they provide compared to large-scale efforts warrants strong support over the next decades. This section presents a broad set of small-scale research projects that could provide key new precision measurements in the areas of electric dipole moments (EDMs) (see Sec.~\ref{sec:rpf3:subsec:edm}), magnetic dipole moments (see Sec.~\ref{sec:rpf3:subsec:mdm}), fermion flavor violation (see Sec.~\ref{sec:flavor_violation}), and tests of spacetime symmetries (see Sec.~\ref{sec:rpf3:subsec:spacetime_symmetry}) and tests with gravity (see Sec.~\ref{sec:rpf3:subsec:gravity_test}).

An area particularly ripe for discovery is the search for EDMs of elementary particles which provide strong constraints on BSM theories like SUSY and, more generally, any with complex phases. Rapid progress in this field is anticipated due to new ideas and techniques from the traditional HEP community as well as adjacent communities in nuclear and AMO physics. For example the storage ring proton EDM experiment~\cite{Alexander:2022rmq}, with its detailed plans innovated off decades of muon magnetic dipole moment experiments, is expected to increase the sensitivity for the proton EDM by four orders of magnitude over the current limit. Likewise, AMO experiments may achieve up to an incredible six orders of magnitude improvement in two decades. EDMs are one of a very small class of experiments capable of probing the PeV scale presently. The discovery of an EDM would set an upper bound on the BSM scale of new physics and serve as a guide to new experiments at the energy frontier. Finally, many EDM experiments (like the srEDM) are capable of sensitive searches of dark matter and dark energy as well.

\section{Science opportunities}
The rare processes and precision experiments covered in our topical group are driven by the search for the unknown physics, dark matter and dark energy. In this section we summarize several white papers covering the relevant experimental proposals and corresponding theory needed to interpret them: EDMs including the proton srEDM experiment, AMO and neutron EDM searches; the theory of hadronic contributions to the muon anomalous magnetic moment and related experimental data; and small, precision experiments to test fundamental spacetime symmetries.

\subsection{Electric Dipole Moments}\label{sec:rpf3:subsec:edm}

The following has been condensed and lightly edited from the executive summary of the white paper entitled ``Electric dipole moments and the search for new physics"~\cite{Alarcon:2022ero}.

\begin{figure}[htb]
    \centering
    \includegraphics[width=0.9\textwidth]{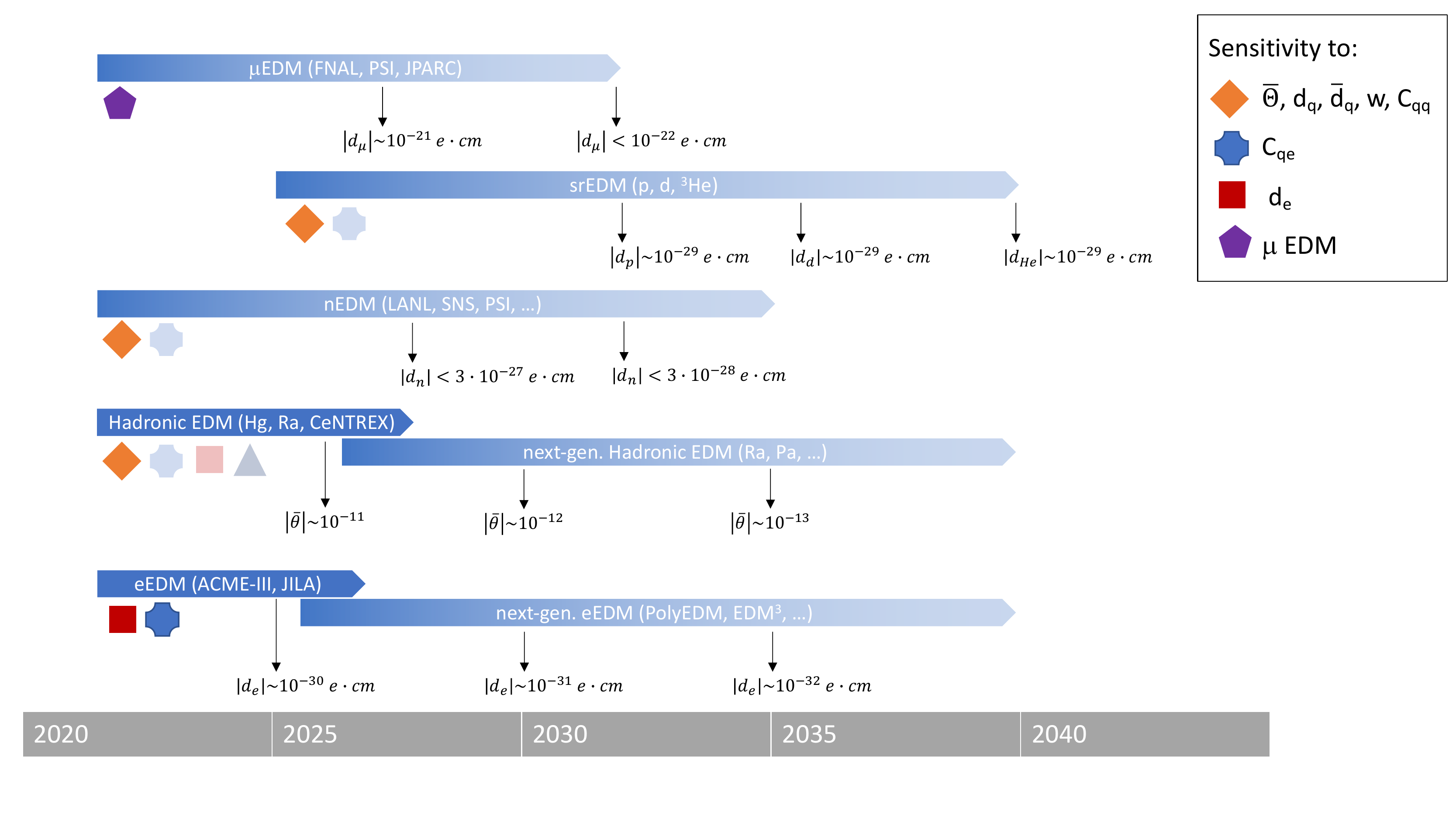}
    \caption{Timelines for the major current and planned EDM searches with their sensitivity to the important parameters of the effective field theory (see Fig.~\ref{fig:edm_scales} for details). Solid (shaded) symbols indicate each experiment's primary (secondary) sensitivities. Measurement goals indicated by the black arrows are based on current plans of the various groups.}
    \label{fig:edm_timelines}
\end{figure}

Observation of a static, ground-state electric dipole moment (EDM) in any experimental system (electron, neutron, proton, atom, molecule) at current or near-future sensitivity would yield exciting new physics. EDMs can probe BSM mass scales well beyond those probed at high energy colliders. To discriminate between the many viable BSM theories, rule out baryogenesis scenarios for the observed matter-antimatter asymmetry of the Universe, or tell whether CP symmetry is spontaneously or explicitly broken in Nature, a well coordinated program of complementary EDM searches in AMO, nuclear, and particle physics experiments is needed (see Fig.~\ref{fig:edm_timelines}). In Ref.~\cite{Alarcon:2022ero} a compelling suite of such experiments and an encompassing theoretical framework are proposed to discover and establish the next fundamental theory of physics.

Searches for fundamental EDMs have a long history, beginning with the neutron EDM (nEDM) beam experiments of Purcell and Ramsey in 1949. Modern experiments use ultracold neutrons (UCN) that can be polarized and stored in room-temperature bottles for hundreds of seconds, leading to very precise measurements. The best measurement of the nEDM stands at $(0.0 \pm 1.1_{\rm stat} \pm 0.2_{\rm sys})\times 10^{-26}\, e\cdot\rm cm$~\cite{Abel:2020pzs} while several UCN experiments are being developed around the world with the goal of reaching slightly above $10^{-27}$ within the next 5--10 years and a few $\times10^{-28}$ in 10--15 years.

Like neutrons, atoms and molecules have been sensitive platforms for precision measurements of symmetry violations for decades, including CP-violation (CPV) through EDMs~\cite{Safronova2018,Chupp2019}.  These experiments now set the best limits on the electron EDM, semileptonic CPV interactions, and quark chromo-EDMs, and are competitive with the nEDM for sensitivity to quark EDMs and $\theta_{QCD}$~\cite{ACME2018,Graner2016}, providing an excellent check on both types of experiments. This class of EDM searches uses techniques from atomic, molecular, and optical (AMO) physics, including very advanced methods for quantum control, to achieve their goals. Improvements in sensitivity of one, two--three, and four--six orders of magnitude appear to be realistic on the few, 5--10, and 15--20 year time scales, respectively, by leveraging major advancements in quantum science techniques and the increasing availability of exotic species with extreme sensitivity.

A whole new class of EDM experiments is possible in storage ring experiments. A ring at BNL or Fermilab, using off-the-shelf technology and a design based on the successful muon \gm experiments can be operational and produce first results for the proton in less than 10 years~\cite{Alexander:2022rmq,Alarcon:2022ero}. The proton storage ring is expected to reach $10^{-29}e\cdot\rm cm$ sensitivity in 10 years, and the same level in an additional five years for the deuteron. The storage ring experiment can also search for wave-like dark matter and energy. EDM storage ring experiments for the electron and muon have been proposed at JLab~\cite{Suleiman:2021whz} and PSI~\cite{Adelmann:2021udj}, respectively. The latter aims for a sensitivity of $6\times10^{-23}e\cdot$cm, about four orders of magnitude better than the limit set by BNL E821. While not a conventional storage ring, the J-PARC g-2/EDM experiment will probe the sensitivity of the muon EDM to $1.5\times10^{-21}e\cdot$cm~\cite{Abe:2019thb}.

Finally, with a beam polarization upgrade at SuperKEKB, the Belle II experiment projects to improve the tau EDM measurement by two orders of magnitude over existing bounds~\cite{Banerjee:2022kfy}.

\begin{figure}[htb]
    \centering
    \includegraphics[width=0.9\textwidth]{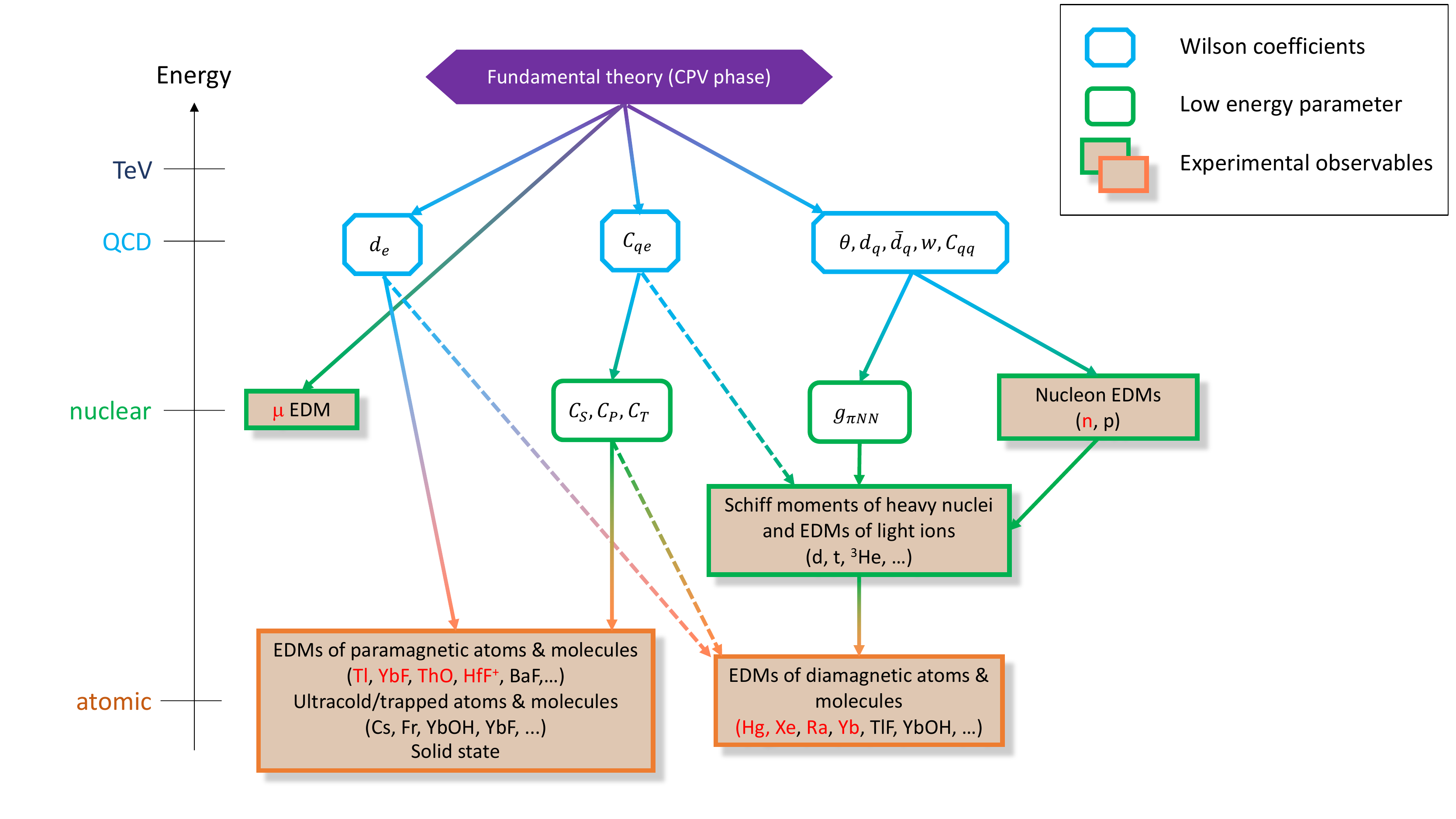}
    \caption{Flowdown diagram from the fundamental physics at high energy scales, to the Wilson coefficients of the effective field theory, low energy parameters, and the experimental CPV observables. Color outlines of the various boxes inidcate the different energy scales. Solid arrows between the boxes indicate strong connection, whereas dashed arrows indicate weaker influence onto the lower lying parameter. Experimental systems shown in red have already been used in EDM searches; those shown in black (as well as many of those in red) are being developed for future searches. This figure was adapted from~\cite{Pospelov2005}.}
    \label{fig:edm_scales}
\end{figure}

Fermion EDMs originate at a high mass scale through new complex CP-violating phases and feed down to lower energy scales in a Standard Model effective theory as summarized in Fig.~\ref{fig:edm_scales}. These elementary particle EDMs then manifest in bound states like the proton and neutron, and even atoms and molecules. At the quark-nucleon level, lattice QCD plays a crucial role in this matching between energy scales. At lower energies still, nucleon chiral perturbation theory, and finally, nuclear and atomic theories are needed. The reverse holds as well: if a nucleon, atomic, or molecular EDM is measured, the underlying BSM physics is tested and can be diagnosed only if the theoretical calculations of the matrix elements reduce their current uncertainties by factors of 2--3. This remarkable effective theory framework, encompassing tens-of-orders of magnitude in energy, is in place today, but the calculations will need substantial effort and computing resources over the next decade. Crucially, the low-energy theory is continually improved and forms a basis for any new high energy models that are invented.

The effective field theory captures the most important effects of the physics at the high-energy scale in terms of a hierarchy of effective `operators’ at a lower energy scale. These operators are organized by a dimension that quantifies the expected power suppression of its effects by the ratio of the low to high scales.  Considering only the electric dipole moments of the light charged leptons, nucleons, nuclei, atoms and molecules, the leading effects at the scale characterizing nonperturbative aspects of QCD, i.e., 300\,MeV--1\,GeV, are the QCD vacuum angle (\(\theta\)), the electric dipole moments of the leptons (\(\mu\) EDM, and \(d_e\)) and the quarks (\(d_q\)), the chromo-electric dipole moments of the quark (\(\tilde d_q\)) and the gluon (\(w\)), and some four-fermi interactions between the quarks and the electron (\(C_{qe}\)), or involving only quarks (\(C_{qq}\)).  The relation between these and the fundamental theory is given in terms of Wilson coefficients that can be calculated perturbatively. At the nuclear scale of a few MeV, these operators, in turn, give rise to the CP-violating interactions of the electron with the nucleons (\(C_{S,P,T}\)), the CPV pion-nucleon coupling \(g_{\pi N N}\), and the EDMs of the nucleons. Obtaining all of these requires the calculation of non-perturbative matrix elements. It is anticipated that only lattice QCD will provide them with the required precision to make full use of the expected experimental constraints in the next decade.  The atomic and nuclear EDMs, at eV to MeV scales, can be expressed in terms of these more fundamental quantities but need quantum mechanical calculations involving nuclear structure as well as the electronic structure of atoms and molecules to relate them.  Figure~\ref{fig:edm_scales} shows these relationships in the effective field theory between the various scales. In the figure solid arrows depict the major determinants at a higher scale for the lower scale physics, whereas the dashed lines indicate less strong influences.

\subsubsection{The proton storage Ring EDM experiment}

The storage ring EDM (srEDM) collaboration has proposed a proton storage ring experiment~\cite{Alexander:2022rmq, PhysRevD.105.032001} based on off-the-shelf technology that can be sited in the 805 m AGS ring tunnel at BNL with an electric field strength of 4.4 MV/m. The expected sensitivity, $10^{-29}e\cdot$cm, could be reached in less than 10 years from the start of construction and represents a three orders of magnitude enhancement for the QCD $\theta$-term EDM over the current limit set by neutron EDM experiments. The experiment has reach to new physics scales as high as 1000 TeV and can probe models of vector dark matter and energy. The various experiments are complemntary, the sensitivity of the proton EDM to CP-violation in the Higgs sector is 30$\times$ that of the electron EDM~\cite{Marciano2020}, for example, due to the smallness of the electron mass. 

The BNL AGS produces $10^{11}$ highly polarized protons per fill. One fill of the EDM ring per twenty minutes is required. As with the anomalous magnetic moment of the muon experiments at CERN, BNL, and FNAL, the key is running at the magic momentum (the proton magic momentum is 0.7 GeV/c.) where the \gm spin process in an electric field is zero. This provided a 2000-fold reduction in systematic uncertainty from CERN II to FNAL E989.

The muon magnetic moment experiments have used magnetic bending and electric focusing. The electric dipole moment experiment will use electric bending with magnetic focusing. The distortion of the closed orbit then automatically compensates spin precession from radial magnetic fields, to first order. With the following symmetries, the systematic uncertainties are estimated to be  below $10^{-29}e\cdot$cm~\cite{Alexander:2022rmq}: storage of clockwise and counter-clockwise bunches, storage of longitudinally polarized bunches with positive and negative helicities and radially polarized bunches, 24-fold symmetric lattice to support the ring, sign change of the focusing/defocusing quadrupoles within 0.1\% of the ideal current
setting per flip, beam planarity better than 0.1 mm, split between counter rotating beams less than 0.01 mm, closed orbit BPM resolution integrated over one second of one $\mu$m for the traditional linear cut
BPM and much better for SQUID BPMs.

With both the BNL E821 and FNAL E989 experiments, new effects were found at the level of sensitivity, but were understood, and mitigated. A similar experience is expected for the pEDM experiment. For the magic momentum in the AGS tunnel, for example, the bending electric field is 4.4MV/m, which is smaller than the Tevatron pbar-p separator electric field. 

The time for construction is estimated at three to five years, with three years of data collection and analysis. 

\subsubsection{Atomic and Molecular EDMs}

The basic idea of AMO-based EDM searches is that CP-violating electromagnetic moments of atomic and molecular constituents (i.e., electrons and nuclei) can be amplified by the extreme internal electromagnetic environment.  The CPV interactions of electrons and nuclei in these ``internal fields'' can be orders of magnitude larger than those directly realizable in laboratory fields.  Due to the presence of both electrons and nuclei, AMO-based searches are sensitive to a wide range of underlying fundamental effects, such as the electron EDM, new CPV nuclear forces, quark-chromo EDMs, $\theta_{QCD}$, and more. The relative sensitivity to these effects depends on the electronic and nuclear structure of the atom or molecule species studied.

It was long understood that the relevant ``internal fields'' in polar molecules can be 3-4 orders of magnitude larger than in atomic systems \cite{Flambaum1985,Sandars1967}. However, only recently have the methods of quantum control used for atoms started to become available for increasingly complex systems such as molecules. New molecule-based experiments using such methods led to a 100-fold improvement in sensitivity to the electron EDM over the past decade. In the meantime, atom-based experiments have achieved steady improvements in sensitivity to CPV signals.

\begin{figure}
    \centering
    \includegraphics[width=0.7\textwidth]{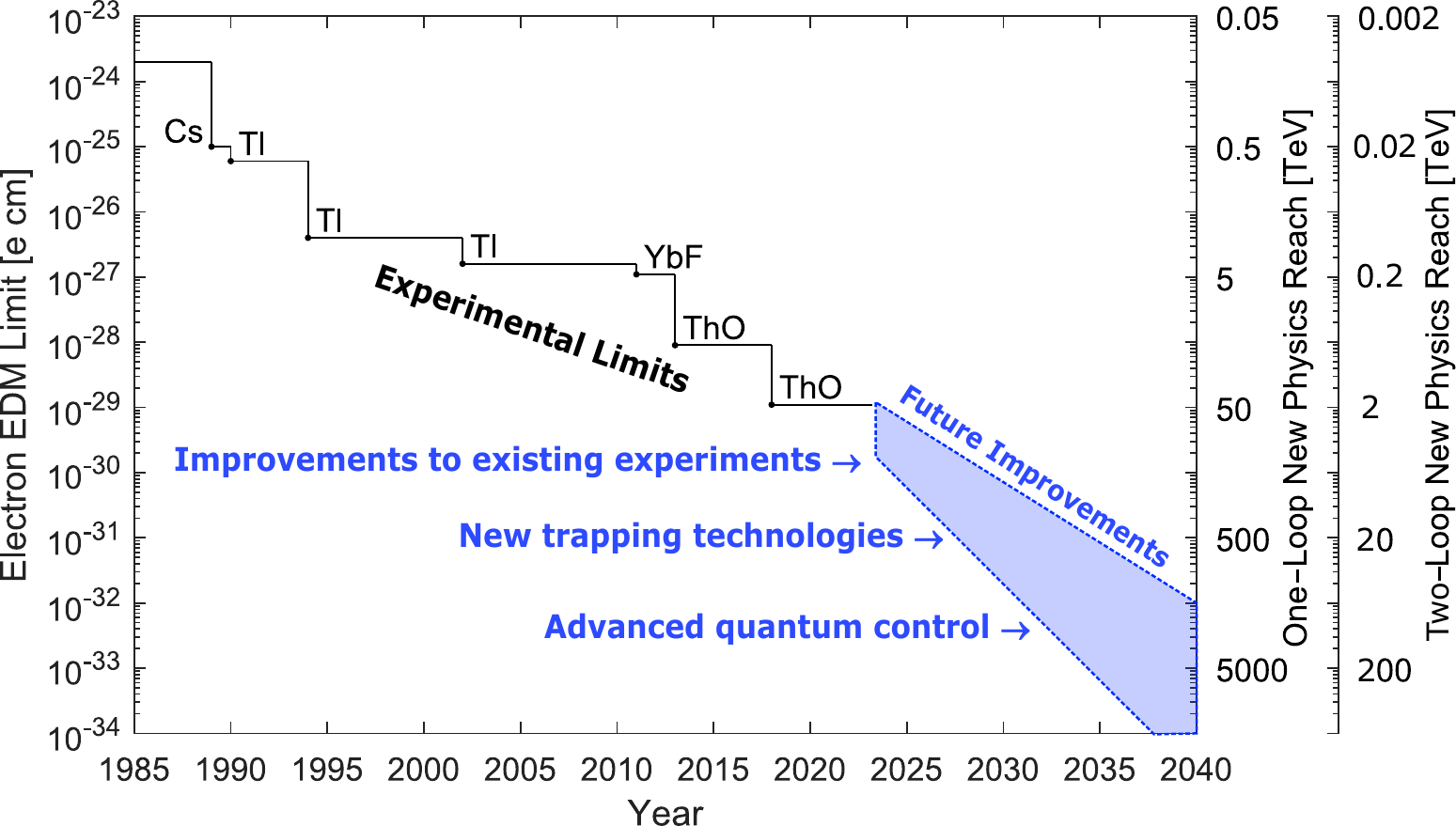}
    \caption{Electron EDM limits versus time, along with new physics reach for one-loop and two-loop effects (see~\cite{Alarcon2022}).  
    The solid line indicates the most sensitive experimental limit, including the species used.  The shaded area indicates potential future improvements discussed in the text.  
    }
    \label{fig:EDMReachVsTime}
\end{figure}

\textit{Improvements of existing experiments.}  Several AMO-based CPV searches have yielded limits in the past decade, including experiments using YbF~\cite{Hudson2011}, HfF$^+$~\cite{Cairncross2017}, and ThO~\cite{Baron2014,ACME2018} molecules for the electron EDM, and using  Hg~\cite{Graner2016}, Ra~\cite{Parker2015,Bishof2016}, and Xe~\cite{Allmendinger2019,Sachdeva2019} atoms for nuclear CPV via nuclear Schiff moments. Most of these experiments are being upgraded to yield improvements, by factors from 2 to 30, within a few years. Several also have longer-term plans for improvements by several orders of magnitude beyond that, by using some of the advances described below. Each of these experiments use considerably different technologies with correspondingly different systematic errors, and use species with different and complementary sensitivities to the underlying physical sources of CPV. Hence, their mutual success is important to advance the overall scientific goals.

\textit{New and extended methods for cooling and trapping.}  
Laser cooling and trapping, along with the quantum control enabled by ultracold temperatures, has been a main driver of quantum science advances with atoms. Extending these techniques and applying them to species with high sensitivity to CPV effects promises to yield orders of magnitude improved sensitivities, by enabling very long spin coherence times and/or enhanced statistics in such systems.  The atomic Ra EDM experiment has provided a preliminary demonstration of the power of this approach~\cite{Parker2015,Bishof2016}. 
Major advances are now being made in the laser cooling and trapping of neutral polar molecules, opening the possibility to also leverage their enhanced sensitivity to CPV effects~\cite{Tarbutt2018a,McCarron2018Review,Moses2017}. Broad advances in molecular ion trapping have resulted in new, successful methods for CPV measurement~\cite{Cairncross2017}, quantum control~\cite{Chou2017}, and the trapping and control of exotic species with extreme sensitivity to CPV effects~\cite{Fan2021,Yu2021}. 
New approaches also are being developed to embed EDM-sensitive species in noble gas matrices, to realize extremely high statistics with potentially long coherence times~\cite{Vutha2018Atoms,Vutha2018PRA,Singh2019,Upadhyay2020}.

\textit{Advances in quantum control.} AMO-based CPV experiments rely on creation of quantum superpositions that are closely analogous to those used in quantum information science (QIS). While the techniques used so far are typically primitive compared to those used in modern QIS experiments, advances in the QIS field are a powerful resource for future dramatic improvements in sensitivity to EDM-like CPV signals~\cite{Cloet2019}. Applying certain types of many-particle entanglement developed for quantum metrology applications---such as spin squeezing---to experiments aimed at detecting CPV signals could provide gains in sensitivity of up to a factor of $\sim\sqrt{N}$ (where $N$ is the number of particles in a single measurement)~\cite{Pezze2018}. Since typically $N\gg 1$ in AMO-based EDM experiments, this could in principle enable improved sensitivity by orders of magnitude; already in atomic clock-like systems a factor of 10 beyond the standard quantum limit has been demonstrated~\cite{Hosten2016}.

\textit{Access to exotic species.} Heavy nuclei with octupole deformations, such as some isotopes of Fr, Ra, Th, Pa, and others, can have sensitivities to CPV enhanced up to a thousand-fold compared to spherical nuclei~\cite{Auerbach1996,Dobaczewski2005,Parker2015,Dobaczewski2018,Flambaum2019Schiff,Flambaum2020Schiff,Flambaum2020SchiffStable}. Combined with molecular enhancements, heavy molecular species with deformed nuclei can be up to $10^7$ times more intrinsically sensitive~\cite{Sushkov1985,Flambaum2019Schiff} to hadronic CPV than the current most sensitive experiment, which uses atomic Hg~\cite{Graner2016}. With such systems, even the ability to trap a single molecule with second-scale coherence would enable probes of new physics at the frontiers of hadronic CP violation~\cite{Isaev2010,Yu2021,Fan2021}.  The short half-lives of these species offer significant challenges, but several short-lived atomic species have been laser-cooled and trapped. Experimental facilities such as GSI/FAIR, FRIB, IGISOL, ISOLDE, and TRIUMF make it possible to perform experiments with short-lived and exotic species; offline sources based on decay of heavier, longer-lived isotopes~\cite{Lu1997,Guckert1998} may also be useful.

\textit{Community effort.} As mentioned above, different atomic/molecular species offer different relative sensitivities to the many potential underlying physical sources of CPV; moreover, each experimental method offers different advantages and different ways to diagnose and control systematic errors.  Interpretation of experimental results in terms of underlying fundamental physics, as well as guidance to identify CPV-sensitive yet experimentally tractable species, requires integrated efforts from a broad range of theorists bridging different traditional communities including atomic, molecular, chemical, nuclear, and particle physics.  To continue the rapid pace of advances in this part of the field and to continue opening new pathways, dedicated efforts are needed to bring these communities together and support a broad portfolio of experimental and theoretical work.

\textit{Summary.} The new and upcoming developments in this field aim to probe a wide range of CPV physics---with signals arising from underlying leptonic, hadronic, and semileptonic CPV interactions---combining one or more of the new approaches discussed above. Improvements by roughly an order of magnitude are anticipated in the next few years, and potentially by several orders of magnitude in the coming 5--10 years. This part of the field is exceptionally fast-moving at present, and offers significant potential for discovery of physics beyond the Standard Model.

\subsubsection{Neutron EDM}

The EDM of the neutron is, at the leading order, sensitive to \(\bar\Theta_{\rm QCD}\), the EDMs of quarks, the chromo-EDMs (cEDMs), and the CP violating (CPV) four-quark interactions~\cite{Shindler:2021bcx}. Unfortunately, the coefficients of these are very poorly known~\cite{Dragos:2019oxn,Bhattacharya:2021lol,Gupta:2018lvp,Bhattacharya:2015esa,Pospelov:2000bw,Lebedev:2004va,Hisano:2012sc,Demir:2002gg,Haisch:2019bml,Weinberg:1989dx}:
\begin{align}
    d_n=&-(1.5\pm0.7)\cdot 10^{-3}\,\bar\Theta_{\rm QCD}\, e\, {\rm fm}\nonumber\\
    &-(0.20\pm0.01)\, d_u + (0.78\pm0.03)\, d_d + (0.0027\pm 0.016)\,d_s \nonumber\\
    &-(0.55\pm0.28)\, \tilde d_u\, e - (1.1\pm0.55\,)\tilde d_d\, e + (50\pm40)\, \tilde d_g\, e\, {\rm Mev}\,,
\end{align}
where all numbers are quoted in the \(\overline{\rm MS}\)-scheme at \(2\ \rm GeV\), \(d\) and \(\tilde d\) denote the EDMs and cEDMs of various particles, the effects of 4-quark CPV interactions, about which even less is known, are not shown.  These large uncertainties dilute the impact of nEDM constraints on BSM physics---in fact, studies of specific models~\cite{Chien:2015xha} show that the matrix elements must be known at the 10--25\% level to reduce the possibility of accidental cancellation between the various sources of CPV.  This level of precision will only be available in the foreseeable future from lattice QCD calculations, which have seen rapid progress in recent years.  Nevertheless, controlling the systematics due to chiral extrapolation to the physical quark mass, light multi-hadron excited state contamination of the nucleon state, the mixing with lower dimensional operators, as well as reducing the statistical fluctuations will need substantial theoretical and computational investment.

The first experiment to search for the nEDM was performed by Purcell and Ramsey in 1949 using a cold neutron beam at the Oak Ridge reactor~\cite{Smith1957}, giving an upper bound of $5\times 10^{-20}$~$e\cdot$cm. Over the last 7 decades, the limit has improved by more than six orders of magnitude, representing significant improvements on the experimental method both to improve the statistical sensitivity and to control the systematic effects at the corresponding level. The current limit ($ |d_n| < 1.8\times 10^{-26}$~$e\cdot$cm, 90\% C.L.) is given by the recent results from PSI~\cite{Abel2020}. This result is dominated by statistical uncertainty. For this reason, there are many efforts worldwide to build UCN sources to host new nEDM experiments: the n2EDM experiment at Paul Scherrer Institute (PSI) in Switzerland~\cite{PSI}, the PanEDM experiment at Institute Laue-Langevin (ILL) in France~\cite{Wurm2019}, LANL nEDM experiment at Los Alamos National Laboratory~\cite{Ito2018}, and the Tucan experiment at TRIUMF in Canada~\cite{Martin2020}. These experiments are based on Ramsey's separated oscillatory method with a comagnetometer and are to be performed at room temperature. They are expected to produce results with sensitivity of $\mathcal{O}(10^{-27})$~$e\cdot$cm in the next $\sim 5$ years. The nEDM@SNS experiment~\cite{Ahmed2019}, which employs a novel approach based on an innovative use of superfluid liquid helium~\cite{GOLUB1994}, is being developed with a sensitivity goal of $\delta d_n = 3\times 10^{-28}$~$e\cdot$cm. Like other nEDM experiments, this experiment provides built-in methods to control the known systematic effects. Finally, an experiment using a pulsed beam of cold neutrons, the Beam EDM experiment~\cite{Piegsa:2013vda,Chanel:2018zga}, is being developed at ILL and is eventually targeted for the European Spallation Source (ESS) in Sweden. Because it has different systematics, it will compliment the UCN experiments.

\subsubsection{Tau EDM\label{tau-emoments}}

\def\CP{\ensuremath{C\!P}\xspace}

The electric dipole moment of the $\tau$ lepton characterizes the T or CP violation properties at the $\gamma\tau^+\tau^-$ vertex. 
The SM predicts $d_\tau$ in the range $10^{-38}-10^{-37}$ $e\cdot\rm{cm}$~\cite{Booth:1993af,Mahanta:1996er,Yamaguchi:2020eub,Yamaguchi:2020dsy}, 
many orders of magnitude below any experimental sensitivity. 
Best results come from a recent Belle study~\cite{Belle:2021ybo}, 
where the squared spin-density matrix of the $\tau$ production vertex is extended to include contributions proportional to the real 
and imaginary parts of the $\tau$ EDM. 
The expectation values of the optimal observable were measured,  yielding
Re$(d_\tau) = (-6.2 \pm 6.3) \times 10^{-18}~{e}\cdot\rm{cm}$ 
and 
Im$(d_\tau) = (-4.0 \pm 3.2) \times 10^{-18}~{e}\cdot\rm{cm}$.
Belle II is uniquely suited to test a large class of new physics models, 
which predict enhanced contributions in EDM of the $\tau$ lepton at observable levels of $10^{-19}$ $e\cdot\rm{cm}$ ~\cite{Bernreuther:1996dr,Huang:1996jr}. 
The proposed beam polarization upgrade at SuperKEKB will further increase experimental sensitivity at Belle~II~\cite{Banerjee:2022kfy}, 
since the uncertainties in modeling the forward-background asymmetry in the detector response are independent of beam polarization 
and will largely cancel. 
Such an increase will enable unambiguous discrimination between the new physics contributions to the $\tau$ EDM at the level of $10^{-20}~e\cdot\rm{cm}$, 
which is two orders of magnitude below any other existing bounds~\cite{Ananthanarayan:1994af,Bernabeu:2006wf, Bernreuther:2021elu}.

\subsubsection{Outlook}

New techniques and technology innovations provide powerful opportunities for elementary particle EDM searches. A proton storage ring experiment can improve the sensitivity by three orders of magnitude within ten years while simultaneously searching for wave-like dark matter and energy. Gains of three to six orders on the AMO side are possible in 10--20 years. UCN experiments are underway or being built that should see sensitivity improve by two orders of magnitude in 10--15 years. Direct lepton EDM searches are expected to make similar gains. Taken together with concomitant increase of precision in the theoretical calulations, the shape of the new physics, should an EDM be discovered, will be sharply constrained. And if an EDM is not discovered, the tight constraints will provide important direction for where to look next.

\subsection{Magnetic Dipole Moments}\label{sec:rpf3:subsec:mdm}
The prototypical precision experiment in our frontier is the measurement of the muon \gm, or anomaly, $a_\mu$. Beginning in the late 1950's at CERN, then moving to the US at BNL, and finally to Fermilab, it has been extremely successful in probing and constraining new physics beyond the Standard Model. Interest in the current experiment, E989, is sky-high, and could continue beyond the expected end of the experiment in a few years, depending on the outcome of on-going theoretical calculations of the hadronic contributions to \gm.

In 2020 the Muon g-2 Theory Initiative released a comprehensive report on the status and value of $a_\mu$ in the Standard Model~\cite{Aoyama:2020ynm}. Taken together with the 2021 Run-1 result from E989~\cite{Muong-2:2021ojo} and the BNL E821 result~\cite{PhysRevD.73.072003}, experiment and theory differ by about 4.2 combined standard deviations. However, at the same time as the Fermilab announcement, the BMW lattice collaboration announced its calculation of the hadronic vacuum polarization (HVP) contribution to $a_\mu$~\cite{Borsanyi:2020mff}, which reduces the discrepancy to about 1.5 standard deviations. 

In the 2020 report, the HVP contribution is taken from data-driven theory, using the $e^+e^-\to$ hadrons cross section and a dispersion relation. The errors from all other collaborations besides BMW are about four times larger than the dispersive ones, which has touched off an intense effort in the lattice community to reduce errors by a factor of four, or more. Experiments at Belle II and future tau-charm factories aim to improve the dispersive result as well. These are the subjects of white papers submitted to our topical group\cite{Colangelo:2022jxc,Huang:2022zqh,BelleII2022}, which we now summarize.

\subsubsection{The muon g-2 in the Standard Model}

\begin{figure}[htb]
    \centering
    \includegraphics[width=\textwidth]{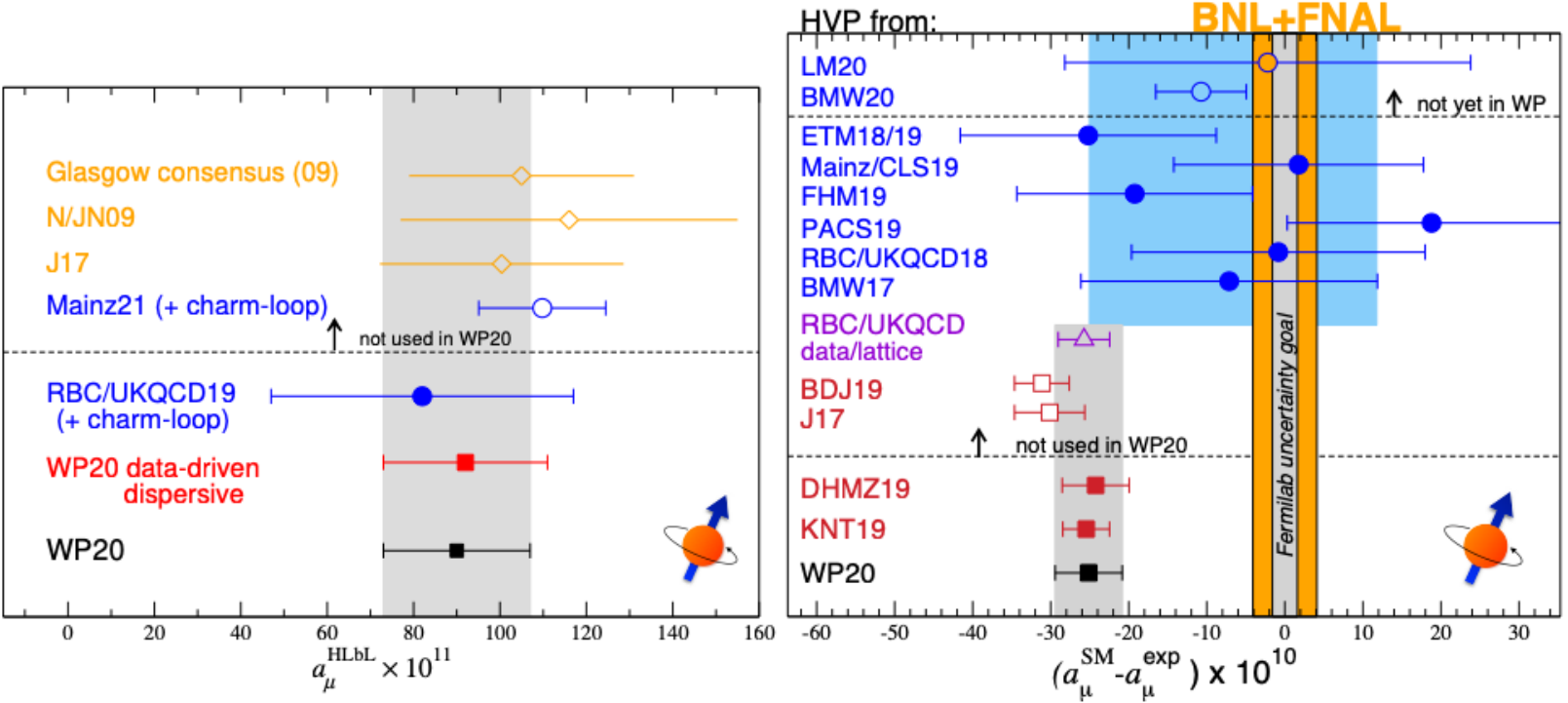}
    \caption{The hadronic contributions to the muon anomaly from hadronic light-by-light scattering (left) and hadronic vacuum polarization (right). Both figures are taken from Ref.~\cite{Colangelo:2022jxc}. Left: Comparison of HLbL evaluations quoted in Ref.~\cite{Aoyama:2020ynm} to earlier estimates (orange) and a recent lattice calculation~\cite{Chao:2021tvp}. Right: Comparison of theoretical predictions of $a_\mu$ with experiment~\cite{Muong-2:2021ojo,PhysRevD.73.072003} (orange band). Each data point represents a different evaluation of the leading-order HVP contribution. Red squares show data-driven results; filled blue circles indicate lattice-QCD calculations that were taken into account in the WP20 lattice average (blue band), while the open ones show results published after the deadline for inclusion in that average; the purple triangle gives a hybrid of the RBC/UKQCD18 and J17 results. The SM prediction of Ref.~\cite{Aoyama:2020ynm} is shown as the black square and gray band and includes only data-driven results.}.
    \label{fig:hadronic contributions}
\end{figure}

The dominant errors in the Standard Model value for the muon \gm~\cite{Colangelo:2022jxc} arise from the leading and next-to-leading order hadronic contributions: the hadronic vacuum polarization (HVP) and hadronic light-by-light (HLbL) scattering. Figure~\ref{fig:hadronic contributions}, reproduced from \cite{Colangelo:2022jxc}, summarizes both. The error on the HVP contribution is about twice the HLbL error, but with more data and improved analysis the HVP may be reduced to where the two are comparable. To reach the expected level of precision of the E989 measurement ($16\times 10^{-11}$), the HVP error ($40\times 10^{-11}$) should be reduced by a factor of three, at least, while the HLbL error ($17\times 10^{-11}$) which is already at about that level should be reduced so the goal on the total error can be reached.

To claim discovery of new physics or test the Standard Model to even better precision it is mandatory to have consistency between dispersive and lattice calculations.
As emphasized in \cite{Colangelo:2022jxc}, the highest priority now is to scrutinize the BMW HVP result \cite{Borsanyi:2020mff} while improving the other lattice results to the same level of precision. At the time of the original Muon g-2 Theory Initiative~\cite{Aoyama:2020ynm} the lattice average for the HVP contribution had an error of about 2.6\% while the later BMW result~\cite{Borsanyi:2020mff} comes in around 0.75\%, resulting in a Standard Model value that is consistent with experiment within 1.5 sigma. However it is only about two sigma away from the data-driven result quoted in Refs. \cite{Aoyama:2020ynm,Colangelo:2022jxc}. It should also be stressed that the BMW error is dominated by systematics, and in particular, the continuum limit extrapolation.

Different lattice groups use different discrete versions of continuum QCD, so results must agree only in the continuum limit, $i.e.$, after extrapolation to zero lattice spacing. While the leading dependence is known analytically, it is challenging numerically to reach the limit. An intermediate step is to chop up the total into smaller chunks according to the Euclidean separation of the two electromagnetic currents comprising the HVP correlation function~\cite{RBC:2018dos}. The ``window" quantities, especially those that avoid both very short and long distances, are already precise enough to detect important differences. For example, for the so-called intermediate window between 0.4 and 1.0 fm separation, Aubin, {\it et al.}, first reported a value significantly above the R-ratio value~\cite{Aubin:2019usy}. The BMW result is 3.7 sigma above the R-ratio (dispersive) result. Other groups find similar discrepancies~\cite{Aubin:2019usy,Lehner:2020crt,Aubin:2022hgm,Ce:2022kxy,Alexandrou:2022amy}, while the RBC/UKQCD result~\cite{RBC:2018dos} is consistent with the R-ratio value. It is expected that by the second half of 2022, several lattice groups will have published window quantities that have errors well below the percent level and that a precise comparison among all of groups will be made.

To tackle the long-distance part of the HVP contribution requires established noise-reduction techniques~\cite{Blum:2012uh,Shintani:2014vja,Giusti:2004yp,DeGrand:2004qw,Neff:2001zr,Bali:2009hu} and recent algorithmic improvements, like two-pion exclusive state reconstruction~\cite{Erben:2019nmx,Bruno:2019nzm}. The two-pion state dominates the long-distance part of the correlation function and can be computed more accurately as a separate contribution. This more accurate ``tail" in turn leads to a more aggressive, improved ``bounding method" where $a_\mu$ is determined by the overlap of strict upper and lower bounds at relatively short distance, and noisy, but exponentially small contributions can be neglected. Again, several groups, including Aubin, $et\,al.$, BMW, ETM, Fermilab Lattice-HPQCD-MILC, Mainz, and RBC/UKQCD are working on next-generation results with sub-percent errors to match BMW that are expected towards the end of 2022 or early 2023. These computations are not cheap, and continued progress to achieve sub-percent and even permille precision requires continued, sustained access to substantial HPC resources.

The dispersive/data-driven error for the HVP contribution is at the 0.5 percent level, but tensions between the most precise data sets from BaBar and KLOE, in the dominant two-pion channel, are a longstanding issue~\cite{Aoyama:2020ynm}. New data
and covariance matrix from SND and BESIII, respectively, are not yet included in the global consensus value. New $2\pi$ results from BaBar, CMD-3, BESIII, and Belle II, are expected soon. If the differences between the BABAR and KLOE experiments can be resolved with the upcoming 2$\pi$ new data and analyses, a precision of 0.3\% seems feasible by 2025~\cite{Colangelo:2022jxc}. 

Then what remains is to resolve possible differences between data-driven and lattice results, which is also likely, as the various calculations, with ever increasing precision, are scrutinized and compared with each other. As implied above, the window quantities can be directly compared to the R-ratio data, so differences between lattice and data-driven values can be isolated to specific distances. Finally, if all differences are resolved, data-driven and lattice results will be combined to yield a result with better precision than either separately.

Figure~\ref{fig:hadronic contributions} shows the current world average for the HLbL scattering contribution to the muon \gm (left panel). Like the HVP, there are both lattice and data-driven results, but unlike the HVP, the two are quite compatible. 

Since the so-called Glasgow consensus value (2009), significant strides have been made in computing the leading pole and pion-loop contributions from data and dispersion relations~\cite{Aoyama:2020ynm} which reduced the error considerably. At the same time, the first complete lattice calculation was also finished, and the two are combined in the world average (see Fig.~\ref{fig:hadronic contributions} with an error of about 18\% which is less than 1/2 the HVP error. It seems the HLbL contribution is now under very good control, and by itself can not explain a difference between the experimental value and the Standard Model. 

On the data-driven side, work remains for the scalar, tensor, and axial-vector channels, as well as short-distance quark-loop contributions. New data from Belle II and a dedicated two-photon program by BES III~\cite{BESIII:2020nme} may help reduce some of the uncertainties~\cite{Accardi:2022oog}, especially for the axial-vectors. Though they are sub-leading, they dominate the uncertainty. For the lattice, a sustained effort is needed to reduce dominant statistical and systematic errors associated with finite volume and non-zero lattice spacing, which in total amount to about twice the data-driven error. The recent Mainz calculation~\cite{Chao:2021tvp} quotes a smaller error but relies on a large chiral extrapolation to reach the physical point. Both the RBC and Mainz groups are improving their calculations and could achieve 10\% precision by 2025, roughly a factor of two improvement over the current data-driven error. 

\subsubsection{New R-ratio measurements}

Both existing and new $e^+e^-$ collider experiments promise significant error reductions in the
data-driven, dispersive, values for $a_\mu$
through more precise measurements of the R-ratio (ratio of $e^+e^-\to$ hadrons to
$e^+e^-\to\mu^+\mu^-$ cross-sections). Analyses of high-statistics $2\pi$ data collected by the BaBar, BESIII, and CMD-3 experiments are ongoing and are expected to provide new, more precise results within the next one-to-two years. In China a symmetric machine~\cite{Huang:2022zqh} running at 2--7 GeV center-of-mass energy is projected to have errors at or below 2\%. Meanwhile Belle II will probe the all-important two-pion channel from threshold to 3 GeV~\cite{Accardi:2022oog}, and compared to Babar and KLOE, ``Belle II will perform these measurements with larger data sets, and
at least comparable systematic uncertainty, to resolve this discrepancy"~\cite{BelleII2022}. The importance of resolving the difference between BaBar and KLOE in the 2$\pi$ channel cannot be overstated: this resolution alone could reduce the total uncertainty by almost a factor of two. Finally new $\tau$ semileptonic decay measurements at Belle II and the super tau-charm factory may provide important additional information to improve the determination of the HVP~\cite{Accardi:2022oog} if better quantification of isospin breaking effects, which may be possible using lattice QCD, is realized. 

\subsubsection{The anomalous magnetic moment of the $\tau$ lepton\label{tau-mmoments}}

Present deviations of the observed magnetic moment of the muon from its SM prediction make tau lepton measurements compelling, as the contribution to the anomalous magnetic moment of a lepton is enhanced by the lepton mass-squared in Minimal Flavor Violation (MFV) scenarios ~\cite{Chivukula:1987fw,Hall:1990ac,Buras:2000dm,DAmbrosio:2002vsn} or generic enhancements in other well motivated models~\cite {Crivellin:2021rbq}. 
The experimental determination of the anomalous magnetic moment  relies on the determination of the cross-section or partial widths for $\tau$-pair production, 
together with spin matrices or angular distributions of the $\tau$-decay products~\cite{Bernabeu:2007rr, Chen:2018cxt}. 
The SM prediction~\cite{Eidelman:2007sb} for the magnetic moment form factor of the $\tau$ lepton is $-2.7 \times 10^{-4}$ at Belle II energies~\cite{Crivellin:2021spu}.  
The current experimental results give ${\mathcal{O}(10^{-2})}$ precision~\cite{DELPHI:2003nah, ATLAS:2022ryk}, 
which is still weaker than the leading term $\frac{\alpha}{2\pi} = 1.6 \times 10^{-3}$ from Schwinger’s famous prediction~\cite{Schwinger:1948iu}.
With 40~ab$^{-1}$ of $e^+e^- \rightarrow \tau^+\tau^-$ data with polarized electron beams, 
requiring both the $\tau^+$ and $\tau^-$ to decay hadronically and assuming a 60\% selection efficiency, 
the statistical error on $a_\tau$ would be $10^{-5}$~\cite{Banerjee:2022kfy}.
As the measurements involve differences in the asymmetries of left-right polarization states of the beam,  
the dominant detector systematic uncertainties cancel. 
Consequently, a polarization-upgraded SuperKEKB  constitutes a promising way to precisely measure $a_\tau$ at Belle~II.
The path towards eventually constraining BSM contributions to $a_\tau$ at the $10^{-6}$ level will require more statistics as well as higher precision measurements of $m_{\tau}$ 
and  $M_{\Upsilon(1S)}$~\cite{Banerjee:2022kfy}, as well as improved modeling of the moments of the $\tau$ lepton in the event generators
{\texttt{KK2F}}~\cite{Jadach:1999vf,Ward:2002qq,Arbuzov:2020coe},
{\texttt{Tauola}}~\cite{Jadach:1993hs,Chrzaszcz:2016fte} and
{\texttt{Photos}}~\cite{Barberio:1993qi,Davidson:2010ew}.

\subsubsection{Outlook}

The muon \gm has sustained intense interest for almost 20 years, from the BNL E821 measurement to the announcement of new results from Fermilab E989. The Standard Model prediction based on data-driven calculations of the HVP contribution shows a 4.2 sigma disagreement. A five sigma or even larger disagreement is possible after E989 releases its final value in a few years. Meanwhile an-almost-as-precise lattice calculation of the HVP contribution, which suggests there may not be a discrepancy, has provided additional motivation to the world-wide lattice community to undertake next-generation calculations with the same, or better, precision. In contrast, both lattice and data-driven HLbL calculations are in good agreement, signaling that any discrepancy between theory and experiment cannot be explained by this notoriously difficult to calculate QCD effect. 

A big hurdle on the experiment side was overcome when E989 released first results that were quite consistent with the BNL experiment E821. While not an issue for concern at this time, the systematics of the storage ring experiments will be tested by a whole new experiment based on cold muons at J-PARC in Japan with first results possibly by 2027. The first phase of the muon g-2/EDM (E34) experiment aims for precision goals of 0.5 ppm for \gm. An upgrade of SuperKEKB to enable polarized beams offers an opportunity to dramatically improve the measurement of the anomaly for tau leptons by several orders-of-magnitude. 

\subsection{Fermion Flavor Universality}\label{sec:flavor_violation}

\subsubsection{Lepton flavor universality with $\tau$ decays at Belle~II}

\def\BF         {{\ensuremath{\cal B}}}
\def\BR         {\BF}
\def\nub        {\ensuremath{\overline{\nu}}}
\def\nueb       {\ensuremath{\nub_e}}
\def\numb       {\ensuremath{\nub_\mu}}
\def\nut        {\ensuremath{\nu_\tau}}
\newcommand{\tauenu}   {\ensuremath{ \tau^{-} \to e^{-} \nueb \nut}}
\newcommand{\taumunu}  {\ensuremath{ \tau^{-} \to \mu^{-} \numb \nut}}
\newcommand{\BRtautomunu}    {\ensuremath{\frac{\BR(\taumunu)}{\BR(\tauenu)} }}
\newcommand{\taupinu}   {\ensuremath{ \tau^{-} \to \pi^{-} \nut}}
\newcommand{\BFtautopinu}    {\ensuremath{\BR(\taupinu)}}
\newcommand{\BFpimutwo}    {\ensuremath{\BR(\pi^- \to \mu^- \numb)}}
\newcommand{\tauknu}   {\ensuremath{ \tau^{-} \to K^{-} \nut}}
\newcommand{\BFtautoknu}    {\ensuremath{\BR(\tauknu)}}
\newcommand{\BFKmutwo}      {\ensuremath{\BR(K^- \to \mu^- \numb)}}

The fundamental assumption that all three leptons have equal coupling to the charged gauge bosons of the electroweak interaction is known as charged-current lepton universality. 
Previous measurements of universality~\cite{HFLAV:2019otj}, expressible in terms of the coupling strength ($g_\ell$) 
of lepton of flavor $\ell$ to the charged gauge boson
of the electroweak interaction, 
are in agreement with the Standard Model (SM)
where $g_\tau = g_\mu = g_e = 1$. 
A broad class of SM extensions violate this assumption, such as two-Higgs-doublet model~\cite{Jung:2010ik} or charged-scalar singlets~\cite{Crivellin:2020klg}, etc. making searches for violation of lepton flavor universality attractive. 
Significant deviations of
this nature are unambiguous signatures of new physics
that provide crucial but complimentary information to
the extended Higgs sector~\cite{Logan:2009uf,Krawczyk:2004na, Loinaz:2002ep}  and other new physics models with leptoquarks~\cite{Dorsner:2009cu}.

Lepton universality is probed by a wide class of experiments, 
such as $\BR(W\to\tau\nu_\tau)/\BR(W\to\mu\nu_\mu)$  
at the LEP~\cite{ALEPH:2005ab} 
and the LHC~\cite{ATLAS:2020xea, CMS:2022mhs},
the $R(D^{(*)})=\BR(B\to D^{(*)}\tau\nu_\tau)/
                \BR(B\to D^{(*)}\ell\nu_\ell)$
~\cite{BaBar:2012obs,LHCb:2017rln,Belle:2019gij}, 
where $\ell$ = $\mu$, $e$, and  
$R(K^{(*)})=\BR(B\to K^{(*)} \mu^+ \mu^-)/
            \BR(B\to K^{(*)} e^+e^-)$
            ~\cite{LHCb:2017avl,LHCb:2019hip,LHCb:2021trn}
anomalies from the B-Factories and LHCb, 
kaon decays from NA62~\cite{NA62:2012lny}, 
pion decays from PIONEER~\cite{PIONEER:2022alm} 
as well as in tau decays from BELLE~II~\cite{Belle2WP}.
New physics that couples primarily to the third generation could be revealed through deviations from the SM in precision universality. 
While LHC measurements of lepton universality via $W^{\pm}$-boson decays are sensitive only to charged currents, tests of lepton flavor universality via $\tau$ leptons offer additional sensitivity to non-SM contributions to weak neutral currents~\cite{Altmannshofer:2016brv, Bryman:2021teu}. 

Measurements in $\tau$ decays typically consist in determining precisely branching-fractions ratios, 
such as $R_\mu \equiv \BRtautomunu$ to test $\mu$-$e$ charged-current lepton universality $g_\mu/g_e$, or $\frac{\BFtautopinu}{\BFpimutwo}$
and $\frac{\BFtautoknu}{\BFKmutwo}$ for $\tau$--$\mu$ charged-current lepton universality $g_\tau/g_\mu$. 
Particle identification is the key experimental challenge, since particle species distinguish between the relevant $\tau$ decay modes. Given the large sample sizes, precision is typically limited by the uncertainties in the corrections needed to match particle-identification efficiencies determined from simulation to those observed in data calibration-samples.    For instance, the most precise measurement of $R_\mu$ from the BaBar experiment~\cite{BaBar:2009lyd} has 0.4\% precision, which propagates to 0.2\% precision on $g_\mu/g_e$,  dominated by the systematic uncertainty associated with lepton identification. 
Early Belle~II data show that lepton-identification uncertainties comparable to those in Ref.~\cite{BaBar:2009lyd} are at reach,
thereby enabling to improve upon the world average values obtained by the HFLAV~\cite{HFLAV:2019otj}.  In this baseline scenario, Belle~II would match the current best results, yielding a significant improvement in global precision obtainable by future HFLAV combinations. In an advanced scenario where an improved understanding of the detector and availability of more abundant and diverse calibration samples reduces lepton-identification uncertainties by a further factor of two, Belle~II would lead to improved precision on the $g_\mu/g_e$ determination~\cite{Belle2WP}.

An important input to lepton-flavor universality tests are measurements of $\tau$ lifetime. 
The global value is dominated by the Belle result based on reconstructing both $\tau$ decays into three charged particles, $\tau(\tau)=(290.17 \pm 0.52 {\rm(stat)} \pm 0.33 {\rm(syst)}) \times 10^{-15}~\rm{s}$~\cite{Belle:2013teo}. 
The size of the Belle~II data set will significantly reduce the statistical uncertainties. The superior control of the vertex-detector alignment, demonstrated in recent charm lifetime measurements~\cite{Belle-II:2021cxx}, will reduce the dominant systematic uncertainty. 
The expected absolute precision of $0.2\times 10^{-15}~\rm{s}$ or better, will further improve the precision of $g_\tau/g_e$.

Belle~II will significantly improve the precision on inputs to lepton-flavor-universality-violating quantities, yielding some of the most stringent constraints on non-SM deviations from charged and neutral current lepton universality~\cite{Belle2WP}. The fact that high precision tests of lepton universality using $\tau$ decays  agree well with the Standard Model are particularly interesting in light of the experimental hints for universality violations observed in semileptonic B decays, the anomalous magnetic moment of the muon, and the Cabbibo angle anomaly~\cite{Bryman:2021teu}.

\subsubsection{Precision electroweak physics with SuperKEKB polarization upgrade\label{sec:ewk}} 

Polarized $e^-$ beams open up new windows of exploration through measurement of the left-right asymmetry $(A_{\it LR})$ of the fundamental electroweak SM  parameter $\sin^2\theta_W$. 
A beam polarization upgrade of the SuperKEKB has been proposed to pursue this unique precision program, along with the rest of the core physics program at the Belle~II experiment. 
This is described in more detail in the  Snowmass 2021 Whitepaper~\cite{Banerjee:2022kfy}. 
$\tau$-lepton polarization measurements would provide precise ($< 0.5\%)$ determinations of beam polarization, 
required for measurements of left-right asymmetry.
With 20~ab$^{-1}$ of beam polarized data, 
the precision on $\sin^2\theta_W$ measured with  
$\tau$s, as well as with muons, electrons, b-quarks and c-quarks, 
will be comparable or better than the current world average.  
Expected precision on $\sin^2\theta_W$ with 
40~ab$^{-1}$ of beam polarized data from SuperKEKB  is 0.0002 at Belle~II.

When $\sin^2\theta_W$ is  measured with multiple fermions, 
these studies will produce neutral-current 
lepton universality measurements of unprecedented precision
as the beam-polarization dependence cancels in the measurements
of the ratio $A^{f_1}_{LR}/A^{f_2}_{LR}$, producing a measurement
of the vector part of the fermion couplings ($g^{f_1}_{V}/g^{f_2}_{V}$).
Contributions from hadronization, which were the dominant component
of the uncertainties of the forward-backward asymmetries of the b- and c-quarks at LEP, are avoided with polarized beams.
For example, $g_V^b/g_V^c$ would be measured with a relative uncertainty below 0.3\% with 20~ab$^{-1}$ of beam polarized data, 
which is an order of magnitude lower than the current uncertainty on this ratio~\cite{ALEPH:2005ab}. In addition, right-handed $b$-quark couplings to the $Z$ can  be experimentally probed with high precision at Belle~II with polarized beams. 
No other experiment, currently running or planned, can perform such precision tests of vector coupling universality in neutral currents~\cite{Banerjee:2022kfy}.

SuperKEKB  will yield the unique possibility of probing “dark forces” that can serve as portals between baryonic matter and dark matter.
SuperKEKB with polarization complements other measurements as it is uniquely sensitive to a parity violating light neutral gauge boson in the dark sector ($Z'$) under various mass and coupling scenarios, including models where $Z'$ couples more to the 3rd generation via mass-dependent couplings. 
For example, a 15 GeV $Z'$ would cause a shift in the measurement of $\sin^2\theta_W$ in the energy region where SuperKEKB with polarized beams may have the best potential for discovery~\cite{Davoudiasl:2015bua}.
With polarized beams, SuperKEKB can also probe parity violating couplings of new heavy particles that couple only to leptons, complementing electroweak studies at the LHC. 
Thus, this ``Chiral Belle" proposal probes parity violation both at very low energies much below the Z-peak, and very high energies~\cite{Banerjee:2022kfy}.

In addition to the precision measurements of the weak mixing angle at 10~GeV, the Chiral Belle physics program with polarized beam also enables the measurement of $(g\!-\!2)_\tau$ at an unprecedented and unrivaled level of precision, as discussed in Section~\ref{tau-mmoments}.
Other  physics uniquely enabled with polarized electron beams includes precision measurements of the tau EDM, as discussed in Section~\ref{tau-emoments}. 
In addition, searches for lepton flavor violation in tau decays and dynamical mass generation hadronization studies will be enhanced with polarized beams~\cite{Banerjee:2022kfy}.

\subsection{Spacetime symmetries}\label{sec:rpf3:subsec:spacetime_symmetry}
Lorentz and CPT symmetry are foundational principles within the boundaries of established high-energy physics as well as key assumptions in many of the Standard Model extensions. In many of these theoretical frameworks, small departures from these symmetries are allowed in particle ground states. Searches for violation of Lorentz and CPT invariance hence poses an important effort within particle physics. Phenomenological and experimental Lorentz- and CPT-symmetry studies therefore fall within the confines of high-energy physics, are critical to the future of the community, and should be intensified.

This section presents a series of ongoing and future efforts that focus on testing spacetime symmetries. Neutron interactions with heavy nuclei are used in the NOPTREX experiment to search for P-odd/T-odd interactions. A multitude of different techniques are testing Lorentz and CPT, both of which are fundamental ingredients for quantum field theories. These efforts include techniques using antihydrogen, clocks, cold neutrons, matter-wave interferometry, muons, resonant cavities, or short-interaction studies and are poised to yield crucial insights into proposed BSM physics. While we give short summaries of the individual efforts below, more details are found in Ref. \cite{Adelberger:2022sve}.

\subsubsection{Time reversal violation searches with NOPTREX}
A search for new sources of time reversal violation is possible using neutron interactions with heavy nuclei. The principle is based on using certain p-wave resonances in the heavy nuclei to search for P-odd and T-odd interactions. Due to the highly excited states involved, the studied neutron interactions offers a qualitatively different environment to electric dipole moment experiments using ground state nucleons and nuclei. The method is based on a ratio measurement rendering it quite insensitive to the resonant state properties and also in principle to final state interaction effects \cite{Gudkov:1990tb, Gudkov:2013nwa, Bowman:2014fca}. Assuming about 4 months of data taking with $^{139}$La at a MW-class short pulse neutron spallation source, an order of magnitude improvement in sensitivity to P-odd and T-odd neutron-nucleus interactions can be achieved \cite{Gudkov:2013nwa, Bowman:2014fca, Bunakov:1982is, Gudkov:1991qg, Beda:2007}, comparable to proposed next-stage neutron EDM searches. Past and ongoing R\&D to polarize $^{139}$La and other heavy nuclei as well as improvements to neutron optics can potentially provide an additional order of magnitude in sensitivity in the future. Conversion of the NOPTREX apparatus into a spin-spin interferometer with recently developed birefringent neutron optical devices could further help to isolate the P-odd/T-odd signal from many possible sources of systematic uncertainties in the measurement.
        
\subsubsection{Lorentz and CPT tests with various low-energy experimental approaches}
Cold antiprotons at CERN's Antiproton Decelerator have been essential for precision measurements with antihydrogen and pave the way for various CPT tests. The ALPHA and ASACUSA experiments will perform antihydrogen hyperfine spectroscopy for direct CPT tests through comparison with the same measurements in hydrogen~\cite{AntiHBook}. The anticipated precision in terms of the mass-antimass difference will even exceed neutral-kaon interferometry, the particle-physics standard for CPT tests.  Other antihydrogen efforts like AEgIS, ALPHA-g, or GBAR at CERN~\cite{AntiHBook} or a proposed future experiment at Fermilab~\cite{AGE}, will study CPT symmetry via the interaction of antimatter with gravity by measuring the free-fall of antihydrogen. Both spectroscopic and gravitational methods will be able to provide qualitatively new Lorentz and CPT tests. 

Another route for probing Lorentz and CPT invariance involve Penning traps,
which enable precision studies of charged particles and their antiparticles. Sidereal time variations of the cyclotron and anomaly frequencies of trapped (anti)particles and the instantaneous comparison of the anomaly frequency between a particle and its antiparticle are common approaches for Lorentz and CPT tests. Future experimental upgrades~\cite{qlr,Portable} 
as well as ongoing phenomenological efforts~\cite{Kostelecky:2021tdf}
would lead to both substantial sensitivity gains
and access to a broader range 
of Lorentz- and CPT-violating effects.

Atomic clocks, atom magnetometers, and other precision spectroscopic experiments provide some of the sharpest Lorentz-violation bounds for protons, neutrons, electrons, and photons~\cite{Kozlov:2018mbp,MB19,SH19,FR17,PG17,PR15,AH14,HL13,MP13,SB11,Cane:2003wp,Bear:2000cd}. Comparisons of at least two transitions in atomic clock can probe Lorentz symmetry violation observed as differences in the clock frequencies. Terrestrial experiments would rely on the Earth motion~\cite{clockKV1,clockKV2}, whereas space-bound experiments access additional forms of Lorentz breaking as they governed by the satellites' motion~\cite{Bluhm:2003un}. Sensitivities will further increase in the next years as clock precision and comparison schemes are improved and new clock technologies are being developed~\cite{Kozlov:2018mbp,Shaniv:2017gad}. 

Cold neutrons provide an indispensable tool in experimental high-energy physics research including Lorentz and CPT tests~\cite{Altarev:2009wd,Babu:2015axa,Ivanov:2019ouz}. Future results from nEDM measurements around the world~\cite{PSI_EDM,Wurm:2019yfj,TRIUMF_EDM,SNS_EDM} 
will give up to about two orders of magnitude improved sensitivities to such tests. The planned NNbar experiment at the ESS~\cite{NNBAR} will provide new sensitivity to neutron-antineutron oscillations including the corresponding distinct class of CPT- and Lorentz-violation searches.

Lorentz violation can also manifest itself as a modification to the interaction of gravity with matter~\cite{akgrav,bailey2006,tasson2009,tasson2011} accessible in superconducting gravimeters or space-based missions which continue to increase their sensitivities~\cite{tasson2017,shao2018,pihan2019}. Matter-wave interferometers are also able to place bounds on Lorentz violation through induced gravitational phenomena~\cite{mueller2008,hohensee2011}. Future, upgraded atom-interferometer using large wave-packet separation or simultaneous multispecies operation promise strong advances in the sensitivity, pushing them to the forefront of the Lorentz tests in the gravitation-matter sector~\cite{space_sep,mom_sep,Schlippert:2014xla,Hartwig:2015iza,Schlippert:2019hzx}.

Muons and muonic systems offer diverse approaches for both Lorentz and CPT tests~\cite{BKLmuon,muonGKV}. 
Spin motion of positive muons in \gm experiments have and will provide access to constraining Lorentz violation in the muon sector~\cite{muong208,muong215,muong219,muong219F}.
Muonic systems such as the theoretically well understood muonium offer another windows to testing Lorentz and CPT invariance~\cite{Muonium01,MuSEUM,Mu-Mass,muonium,Antognini:2018nhb}.

Sensitivity to Lorentz violation in the photon sector is achieved with electromagnetic resonant cavities where one typically compares resonant frequencies of two differently orientated cavities~\cite{microwave1,microwave2,microwave3,microwave4,microwave5,microwave6,nonminmicrowave,optical1a,optical1b,optical1c,optical2a,optical2b,optical2c,optical2d,rings1a,rings1b,rings2,nonminrings1,nonminrings2,acoustic1,acoustic2,ligo,nonminmicrowave,nonminrings1,nonminrings2}. The trend of orders of magnitude improved sensitivities in various type of cavities is expected to continue in future experiments.

Tests of the inverse-square law and searches for novel interactions can often also probe spacetime symmetries due to their geometrical setup of test bodies~\cite{SRG_Theo}. Some of the best constraints in the gravity sector are obtained this way~\cite{SRG_Exp1,SRG_Exp2} and future experimental efforts are leading to improved sensitivities~\cite{Chen:2017bru}. New techniques like a spin-polarized torsion pendulum can further push the frontier on setting better limits on spatial-anisotropy coefficients~\cite{bluhm-00a,ni-03,heckel-06,heckel-08,Luo:2020gjh,Zhu:2018mrf,Lee:2020zjt}.

\subsection{Precision tests with gravity\label{sec:rpf3:subsec:gravity_test}}
Gravity and its full understanding remains a vivid and important field of research. Efforts in this area include precision gravitational tests with antimatter and especially the sign of the force between matter and antimatter. A repulsion predicted in an ``antigravity" scenario would explain several cosmological observations without the need for cosmic inflation. Other important active fields of research in the gravitational sector include the probing of general relativity and the quantum nature of gravity, or searches for short-range corrections to the laws of gravity. The efforts in this section are detailed more in Ref.~\cite{Adelberger:2022sve}.

\subsubsection{Antimatter gravity tests with MAGE}
Muonium---the bound state of an antimuon and electron---provides access to direct antimatter gravity tests with a theoretical interpretation that is free of hadronic effects as in $\overline{\mathrm{H}}$ \cite{Kirch:2014mna, Kirch:2007wa}. The proposed concept uses a high-quality muonium beam in a Mach-Zehnder-type interferometer to measure a small, gravity-induced phase shift, with soft x-rays providing a needed calibration \cite{MAGE:2018wxk}. Ongoing R\&D for the high-quality muonium beam aims at reducing the 6D emittance of a surface muon beam by ten orders of magnitude in order to stop it in a thin superfluid helium (SFHe) layer for muonium formation. A one month data taking at PSI, Switzerland, could lead to a 5$\sigma$ determination of the sign of $\overline{g}$ (i.e., a 40\% measurement of $\overline{g}$) by the LEMING experiment. An alternative approach using a much thicker SFHe layer can avoid the need for the low intensity cooled muon beam envisioned at PSI. It could be developed in parallel at Fermilab and initially lead to a 10\% measurement of $\overline{g}$. This could be improved to 1\% or even better at a future Fermilab facility that would provide competitive muonium beams in the PIP-II era. The MAGE experiment could yield the first gravitational measurement of leptonic matter, 2nd-generation matter, and possibly of antimatter \cite{MAGE:2018wxk}. These developments are also synergistic with possible efforts for precision measurements of the muonium spectrum and searches for muonium-antimuonium oscillation. The needed cryogenic stopping-target R\&D could start almost immediately at the Fermilab Linac. The required slow-muonium facility would consist of a muon production target (or targets) illuminated by a proton beam from a linac: either the existing 400-MeV Linac, or the PIP-II 800 MeV one. R\&D at the 400 MeV Linac can use the Irradiation Test Area (ITA), where an effort is already in progress \cite{Johnstone_private} to produce low-energy muons for other applications. The needed antimuons would be deflected into the vertical before entering a small cryostat, cooled by a dilution refrigerator to sub-Kelvin temperatures, in which a layer of superfluid helium would serve as an efficient $\mu^+$-to-muonium (Mu) converter, as well as deflecting the Mu beam into the horizontal, thus producing a quasi-monoenergetic, quasi-parallel muonium beam in vacuum. This beam would enable (especially at PIP-II) world-leading sensitivities for muonium gravity, muonium spectroscopy, and Mu—$\overline{\rm Mu}$ oscillation experiments.

\subsubsection{Gravitational effects on CPV}
An indirect measurement of antimatter gravity with kaons is proposed via the measurement of the magnitude of CP violation in different gravitational field strengths \cite{PIACENTINO2016162, Piacentino_2017, Piacentino_2019, Piacentino:2021xzt, LOI5, Adelberger:2022sve}. The approach is motivated as many Standard Model extensions imply a large CP violation and antigravity. Gravity-generated CP violation \cite{PhysRev.121.311} could also help to understand the missing antimatter in the universe. The Earth's gravitational field strength at the surface and the long enough mixing time of the $K^0-\overline{K}^0$ system would yield a separation of the antimatter components of the $K$ meson \cite{Chardin:1992mb}. Assuming linearity of the CP violation with gravitational field strength, the same effect would be $\sim$97\% smaller on the moon. The proposed concept is based on a comparison of the ratio of $K^{}_L$ decay into two and three pions in either a low orbit near the Earth's surface or on the moon. In the absence of a particle accelerator in these lower gravity environments, the production of $K^{}_L$ would leverage incident cosmic ray protons. A possible discovery of the CP violation dependence on the gravitational strength could then motivate a dedicated laboratory in space to perform similar measurements in greater detail. 

\subsubsection{Tests of general relativity with a $^{229}$Th nuclear clock}
Precision time-keeping plays an important role in many areas of physics. Further advances in the precision of clocks has the potential to reveal new physics through tests of the constancy of fundamental constants or tests of general relativity. A nuclear transition with an energy low enough for laser excitation in $^{229}$Th is a promising candidate for a novel nuclear clock that could push the boundaries of the clock precision by 2--3 orders of magnitude. Much of the ongoing R\&D to better understand properties of $^{229}$Th like the half-life of its meta-stable state $^{229}$mTh or the exact transition energy needs to continue before such a nuclear clock could even be realized that would open the pathway to ultra-precise new tests in the gravitational sector.

\subsubsection{Mechanical tests of the quantum-gravity interface}
Tests of the quantum nature of gravity would require experiments at the Planck energy scale. Low energy probes offer an alternative that is accessible today \cite{Carney_2019}. One such possibility is to attempt the entanglement of two masses that are prepared in a quantum state of their motion. Two classes of experiments can probe such gravitational quantum entanglement: i) interferometric tests that rely on preparing masses in a quantum superposition of their positions \cite{Feyn57,Bose:2017,Marletto:2017}, which would dramatically decohere when exposed to classical gravity. New interferometric experiments with levitated nano-particles are planned for the next decade. As they are ultimately limited by the free-fall-time on Earth, they provide the test-bed for subsequent space-borne setups. ii) Non-interferometric tests that hope to precisely account for and measure the subtle effect of gravitational entanglement \cite{Diosi1989,penrose}. Such tests using mechanical oscillators prepared in quantum states are poised to enter the regime where gravity can be sourced and sensed using quantum objects. This state of advance is largely due to the recent progress in understanding the operating principles and limits of quantum-noise-limited displacement measurement and control of mechanical motion at the quantum level. A new generation of table-top experiments are being planned to set stringent bounds on gravity’s ability to mediate entanglement \cite{PedernalesPRL2020, WoodPRA2022, vandeKampPRA2020, TorosPRR2021, Amit_2019, Margalit_2021, Marshman_2021, Henkel_2021,Japha_2022}. These new experiments of both types share the need to understand and develop experimental techniques of broader impact such as low-environmental noise, mitigation of thermodynamic noises (for example via low-noise cryogenics, materials science, and engineering), and shaping of quantum noises (for example, via quantum-enhanced metrology and control). A dedicated low-noise underground user facility and the use of freely-falling platforms would be beneficial for advancing progress in this area.

\subsubsection{Searches for short-range corrections to gravity}
To help understand the more than 16 orders of separation between the apparent energy scale of quantum gravity and that of the Standard Model (electro-weak scale), investigations of the behavior of gravity at sub-millimeter distances has been theoretically motivated \cite{add,GiudiceDimopoulos}. Such low-energy experiments are challenged by the weakness of gravity and hence require ultra-precise measurements. A wide set of experimental approaches is available, among which are torsion pendulums, slow neutrons, or optically levitated sensors.

Torsion pendulums remain one of the promising paths forward in studying exotic short-range gravity \cite{Kapner:2007,Lee2020}, equivalence-principle violations \cite{wagnerEPV,Shaw2022}, or novel spin-dependent interactions \cite{Terrano:2015}. Both vibrations and time-varying gradients from the environment are often the limiting factors. Future progress with this powerful tool could hence be driven by the development of a suitable underground facility that would strongly suppress such environmental backgrounds.

Slow neutrons offer another versatile tool due to its special properties (like electric neutrality with small magnetic moments and electric polarizability) that renders them insensitive to many electromagnetic backgrounds. This makes them very suitable for unique types of precision measurements \cite{Leeb92, Bae07, Ser09, Ig09, Pie12, Yan13, Lehnert14, Jen14, Lemmel2015, Li2016neutron, Lehnert2017, Haddock2018b, Cronenberg2018}. The ability of slow neutrons to penetrate and interact in matter with little decoherence allows for interferometric measurements of large phase shifts of their quantum motion amplitudes. These advantages have been exploited in many searches of new weakly coupled interactions and particles of various types. Furthermore, the dominant s-wave interaction for slow neutrons is experimentally well measured and hence making their use as a probe insensitive to strong nucleon-nucleus interactions. As such, slow neutrons are also a good probe for short-range modifications of gravity at the length scales of 100 microns corresponding to the scale set by dark energy densities. Ongoing and future prospects using ultracold neutrons show great potential for continued progress as they are often not yet limited by the statistical accuracy available at current neutron facilities. Further options will arise from neutrons at higher energies that undergo a nuclear resonance reaction which provide an amplification of the small effects by the much prolonged interaction time compared to low energy scattering. 

Optically levitated dielectric objects in ultra-high vacuum are well suited for high precision sensing as they are mostly decoupled from their environment \cite{GeraciMoore2020}.  For example, optically-trapped nanospheres are promising candidates to sense very feeble forces, accelerations, torques etc. \cite{Ranjit:2016, Moore2017,novotnydrop, Li2016, Li2018, Novotny2018, Moore2018}. Levitated objects also serve as sensitive probes for millicharged particles \cite{Moore:2014}, gravitational waves \cite{GWprl}, or dark matter \cite{Monteiro:2020wcb}. Trapped, levitated  spheres can be used as test masses to probe for non-Newtonian gravity-like and other forces \cite{geraci2010,Gratta2022}. Advances in sensitivity made possible by pushing the sensitivity of these sensors into the quantum regime along with improved understanding and mitigation of systematic effects due to background electromagnetic interactions such as the Casimir effect and patch potentials will enable several orders of magnitude of improvement in the search for new physics beyond the Standard model. 

\section{Sensors}\label{sec:techniques}
Many of the precision tests described in Sections~\ref{sec:rpf3:subsec:edm}--\ref{sec:rpf3:subsec:gravity_test} often require new detector technologies to probe the unexplored regions of some parameter space and increase sensitivity to symmetry violations and new physics. In more and more cases in the recent past, experimental efforts and their detection systems reach the regime in which the laws of quantum mechanics become relevant. New quantum sensors have emerged and are actively being developed to use such quantum effects to their advantage and even achieve enhancements in the detection sensitivity. Ongoing efforts and current opportunities are highlighted in a subset of sensor types, such as interferometry, optomechanical sensors, clocks and trapped atoms/molecules. Many of these specific tools have high relevance in particle and high energy physics and other communities as can be seen from some of the explicit examples summarized in Sections~\ref{sec:rpf3:subsec:spacetime_symmetry} and \ref{sec:rpf3:subsec:gravity_test}. More details can be found in \cite{Carney:2022rlu} as only a short summary is given here.

\subsection{Atom Interferometers}\label{subsec:interferometers}
Atom interferometers have acquired a growing number of applications in fundamental physics. Applications range for example from gravitational wave detection, to the dark matter and dark sector, precision determination of fundamental constants (like the fine structure constant $\alpha$), and precise tests of the SM. One specific growing area is the pursuit of long-baseline atomic sensors for GW detection. Significant progress on their sensitivity is expected with the next generation which are proposed for both terrestrial and space-based setups. 

Several types of interferometers exist that have their own applications. Light-pulse atom interferometry uses laser pulses to coherently split, redirect, and recombine matter waves. The gradiometer configuration uses two identical atom interferometers to run simultaneously at each end of a long-baseline. Comparison of both signals enables cancellation of common mode noise sources, which will allow for the possibility of single baseline GW detection. To use an atom interferometer for GW detection, two freely falling atoms on each end would act simultaneously as inertial reference and precise clock. Laser light traveling between the two atom ensembles drives an atomic transition and encodes the light travel time as a phase shift between the two systems. Differential phase measurements are then sensitive to baseline changes induced by a passing GW. Atom interferometers for GW detection are promising to access a frequency gap between LIGO/Virgo/KARGA and LISA. This frequency band of about 30\,mHz to 3\,Hz offers a variety of compelling physics questions to be addressed. 

In addition to GW detection, long-baseline interferometers can also be sensitive to dark matter and new forces. For example, dark matter can affect fundamental constants and with it the energy levels of the involved quantum states in the interferometers. Dual-species interferometers operated with different isotopes are sensitive to acceleration changes that could be caused by dark matter. Finally, a comparison of atomic sensors with different atomic species could also reveal the existence of new forces.

A prominent example for smaller atom interferometers is the application for the precise determination of fundamental constants, such as the fine structure constant $\alpha$, a measurement highly motivated by the ongoing determination of the muon's magnetic anomaly. Other applications using optical cavities can place stringent limits on dark energy candidates. 

The selected examples showcase the versatility of atomic interferometers and their relevance to high energy physics questions. Future improvements to the technique will lead to significantly more sensitive detectors for precision measurements or new limits on BSM physics.

\subsection{Optomechanical}\label{subsec:optomechanical}
Mechanical sensors that are read out by optical or microwave light have undergone significant developments and many of them are now operated in the quantum regime. These mechanical devices are well suited to look for signals acting coherently over the typical size of the mechanical system. Mechanical sensors also come in a wide variety of sizes, operational frequencies, and other configuration parameters.

Besides their prominent use in GW detection and precision measurements in metrology, they have a widening use in particle and high energy physics such as ultra-light and ultra-heavy dark matter particle searches, neutrino detection, or fifth-force modifications to Newton's law. Various types of ongoing and future experiment include the use of for example torsion balances, optomechanical interferometers, resonant mass detectors, nanomechanical systems, or levitated particles.

Further progress in the next years is expected from several opportunities. One such R\&D example is the goal to pass the so-called Standard Quantum Limit to achieve advanced quantum techniques. Many of the improvement opportunities are in need of  theoretical effort to be able to implement advanced sensors.

\subsection{Clocks}\label{subsec:clocks}
Optical clocks and their precision have improved steadily and they play an important role in many applications like precision tests of the constancy of fundamental constants, dark matter searches, or tests of Lorentz invariance and general relativity.

Atomic clocks are based on transitions between hyperfine substates or electronic levels. Comparison of the frequencies of two optical clocks are sensitive to variations in $\alpha$ and with it to SM dark matter couplings. The sensitivities can experience enhancement factors depending on the electronic structure of the clock and tends to increase for atoms with heavier nuclei. Future development of novel clocks with larger enhancement factors can hence provide stronger limits on dark matter. Several R\&D directions are being pursued to achieve this goal. One examples is the development of clock networks with new precision levels, which will yield better comparisons with spatially separated clocks. Another focus area is the increase in clock performance to surpass the standard quantum limit by introducing highly entangled quantum states. Use of highly charged ions or the development of a nuclear clock will provide significant increase in the sensitivity to $\alpha$ through large enhancement factors and represent other areas of ongoing clock R\&D. Finally, molecular clocks with their rotational and vibrational degrees of freedom offer direct sensitivity to $\frac{m^{}_p}{m^{}_e}$ and its variation. 

\subsection{Trapped atoms and ions}\label{subsec:traps}
Radioactive atoms and molecules offer extreme nuclear charge, mass, and deformations advantageous for studying nuclear structure, symmetry violations, and precision measurements. Sensitivity to symmetry violations typically scales with the number of the nucleus as $Z^{2-5}$ favoring heavy, radioactive nuclei. Other special properties can bring further enhancements making certain species like Fr, Ra, Ac, Th, or Pa attractive for CP-violation studies. Certain unstable isotopes are also candidates for improved clocks or allow through precise studies of the nuclear decay to search for sterile neutrinos. As these heavy, radioactive species of interest are difficult to produce in large quantities and decay quickly, traps have been critical to make efficient use of them. R\&D for further improvement of ion and neutral traps offers an important opportunity to make significant progress in the physics measurements.

\section{Summary}

The HEP community has entered an era of unprecedented precision experiments. While relatively small in size and cost compared to their energy frontier cousins, they are large in reach and discovery potential. The traditional storage ring experiments for the muon \gm coupled with equally precise theory calculations have put us on the cusp of discovery, while new experiments and improvements in lattice QCD calculations offer exciting possibilities for EDMs. The proton storage ring experiment is estimated to improve sensitivity by over three-orders-of-magnitude and will also search for wave-like dark matter and energy. Meanwhile adjacent nuclear and AMO physics communities will provide critical complementary searches, the latter with up to six orders-of-magnitude improvements in the next 10--15 years. Together these efforts will provide a sharply focused framework for new physics, or if no EDMs are found, will point our field in important new directions.

\bibliographystyle{JHEP}
\bibliography{RARE/RF03/myreferences}

\end{document}